\documentclass[12pt]{article}  
\usepackage{graphicx}
\usepackage[centertags]{amsmath}
\usepackage[english]{babel}
\usepackage{amsthm}
\usepackage{mathtools}
\usepackage{fancyhdr}  
\usepackage{setspace}   
\usepackage{caption} 
\usepackage{multirow}
\usepackage{amsfonts}
\usepackage{amssymb}
\usepackage{natbib}
\usepackage{multirow}
\usepackage{algorithm}
\usepackage[noend]{algpseudocode}

\title{ \Large \textbf{STAD Research Report 01/2016} \\ \vspace{10mm} Dynamic recursive tree-based partitioning for malignant melanoma identification in skin lesion dermoscopic images. \\   }

\author{Roberta Siciliano*, Massimo Aria**, \\
Antonio D'Ambrosio** and Valentina Cozza***\\
\\\small *Department of Industrial Engineering, 
\\\small University of Naples Federico II\\ 
\small roberta@unina.it\\
\\\small **Department of Economics and Statistics,
\\\small University of Naples Federico II\\
\small aria@unina.it, antdambr@unina.it\\
\\\small ***Department of Law,
\\\small University of Naples Parthenope\\
\small valentina.cozza@unipathenope.it\\}

\date{March 23, 2015}

\begin{document}
\maketitle
\newpage


\noindent \textbf{Abstract}.
In this paper, multivalued data or multiple values variables are defined. They are typical when there is some intrinsic uncertainty in data production, as the result of imprecise measuring instruments, such as in image recognition, in human judgments and so on.\\
\noindent So far, contributions in symbolic data analysis literature provide data preprocessing criteria allowing for the use of standard methods such as factorial analysis, clustering, discriminant analysis, tree-based methods. As an alternative, this paper introduces a methodology for supervised classification, the so-called Dynamic CLASSification TREE (D-CLASS TREE), dealing simultaneously with both standard and multivalued data as well. For that, an innovative partitioning criterion with a tree-growing algorithm will be defined. Main result is a dynamic tree structure characterized by the simultaneous presence of binary and ternary partitions. A real world case study will be considered to show the advantages of the proposed methodology and main issues of the interpretation of the final results. A comparative study with other approaches dealing with the same types of data will be also shown. D-CLASS TREE outperforms its competitors in terms of accuracy, which is a fundamental aspect for predictive learning.
\\


\section{Introduction}
This paper was designed to deal with a real problem of statistical analysis in medical field. The genesis was the analysis of the database of the Department of Dermatology of the Second University of Naples consisting of $220$ skin lesion dermoscopic images, for which a histological diagnosis is available such to divide the images into two classes: $86$ images are relative to malignant melanoma and $134$ of these lesions are classified as benign lesions. Skin lesion dermoscopic images are acquired using a charge-coupled devise camera connected to an epiluminescence microscopy. Descriptors of these images include $11$ point values variables, $6$ intervals data variables and $17$ histograms data descriptors. Thus the database is built up by both standard data as well as non-standard data. The latter are multivalued data, also known as symbolic data, consisting of both interval and histogram data, thus either an interval or a distribution rather a single value. For more details on symbolic data, see Section 2 and the references therein included. This database has been analyzed by a three-steps methodology provided by \cite*{cozza} which considers a dynamic clustering approach. Although this approach has been suitably defined to deal with symbolic data, it does consider an unsupervised method rather than a supervised approach as it should be the case when data include a response variable.\\
Key idea is to innovate tree-based methodology for supervised classification such as CART approach \citep*{cart, statlearn} in order to deal with both standard and multivalued data as well. In literature, proposals of tree-based methods in symbolic data analysis just consider interval data and adopt a suitable data pre-processing to apply the standard splitting criterion with the related partitioning algorithm. As an alternative, this paper provides an innovative partitioning criterion with a tree-growing algorithm. Main result is a dynamic tree structure with the simultaneous presence of binary and ternary partitions. The so-called Dynamic CLASSification TREE (D-CLASS TREE) methodology can be fruitfully considered to deal with different types of data (point data, interval data, histogram data). The results of a real world case study will be considered to show the advantages of the proposed methodology and main issues of the interpretation of the final results. A comparative study with other approaches dealing with the same types of data will be also shown. D-CLASS TREE will be demonstrated to be stable and more accurate, outperforming its competitors in terms of predictive learning.

\subsection{Tree-based methods}

Data can be hierarchically organized in a connected and oriented graph, the so-called tree, characterized by a set of linked nodes, in which any two nodes are connected by exactly one simple path, the starting-node is the \emph{root} and the end-nodes are the \emph{leaves}. Two properties are satisfied: the \emph{shape property}, where each node has a fixed number $r$ of child nodes (for $r=2$ it is assumed a binary tree); the \emph{heap property}, where each node is greater than or equal to each of its children according to some comparison predicate which is fixed for the entire data structure \citep*{sicilianotrees}.\\
Tree structures can be fruitfully considered in both supervised classification as well as non parametric regression. The standard data set consists of a sample of $n$ objects on which are measured a set of predictors (of numerical or categorical type) and a response variable, either categorical (in classification trees) or numerical (in regression trees). In supervised classification there is a prior class assignment of the target or response variable and the predictors play the role to generate the set of candidate partitioning variables to be considered for partitioning the objects of a given node into $r$ subgroups.\\
The sample of objects is recursively partitioned into $r$ subgroups such to reduce the impurity of the response variable within each subgroup as measured by the Gini's diversity index or any suitable entropy measure. Partitioning of the objects is determined by the best partition of any predictor's cateogies into $r$ subgroups. Thus, any predictor plays a role to generate the candidate partitioning variables defined as all possible partitions of the predictor's categories into $r$ subgroups such to induce the partition of the objects. The number of partitioning variables of each predictor depends on the number of distinct categories assumed by the predictor, which can be very high if the domain is real. In case of binary trees, the partitioning variables are known as splitting variables.\\
Leaves of a tree are terminal nodes, thus nodes which are not further partitioned as soon as the impurity or the node size is lower a given value. To each leave of the tree is assigned a response class which provides the posterior class assignment of the object falling into that node. It is possible to interpret this class assignment analyzing the partitioning variables determining the path from the starting node until that leaf.\\
Tree-based models can be also considered for prediction of a new object for that only the predictors' measurements are known. This object can slide down the tree until falling into a terminal node where a response class will be assigned. The quality of this prediction can be evaluated in terms of the misclassification rate estimates considering either the learning or training sample, or the test sample, alternatively cross-validation.\\
There is the classical trade-off between bias and variance in the final misclassification estimate, for that it is necessary to identify the proper tree model complexity. Large tree structures are inaccurate because of a large variance (too much sensitivity  sample) whereas tree-models with too few leaves are inaccurate because of a large bias (not enough flexibility). In the pioneer work, CART methodology \citep*{cart} suggests to grow the maximal expanded tree (fixing a small percentage of objects within each leave), then find a sequence of nested pruned trees (cutting off at each step the weakest link in terms of cost-complexity parameter which takes into account the size reduction and the misclassification error rate increase), finally select the best pruned tree on the basis of the cross-validation or test sample estimate of the misclassification error rate \citep*{cappelli}.

\subsection{Accuracy of decision tree-based rules}

Main focus of recent literature is to outperform the decision/prediction rule of CART methodology in terms of accuracy such to answer the bias-variance dilemma with alternative solutions. Enhancements are provided by ensemble methods, random forest, evolutionary programming. All these approaches do not provide one tree structure for prediction denying the interpretability advantage of the tree graph to describe the hierarchical dependence relationships. The final assignment of a new object is induced by a suitable combination of tree structures. Ensemble methods are learning algorithms that develop a population of simple models (i.e. trees), called weak learners, from the perturbed training set combining them to form a composite predictor, which is generally more accurate than the single trees whence it is formed by. A new observation is classified on the basis of a majority vote of their predictions. There exists several ways to build ensembles \citep*{dietterich}, the most popular ensemble methods, such as Bagging \citep*{bagging}, Boosting \citep*{boosting} and Random Forest \citep*{forest}, work by manipulating the training examples through re-sampling methods. All of these algorithms aggregate the object decisions by voting, but none of these ensemble methods allows to preserve the final tree-structure. If we are interested in the accuracy of the prediction then we can use an ensemble classifier/regressor because it is generally more accurate than a single decision tree (see in example \citet*{dambrosio, borgoni} in the framework of missing data imputation, or \citet*{bashir} in the field of breast cancer diagnosis). In this case the interpretation of the tree-structure is irreparably lost because the aggregation process bars the construction of a unique prediction tree structure.

\section{Multiple valued data description}

\begin{figure}[h!]
	\centering
		\includegraphics[width=1\textwidth]{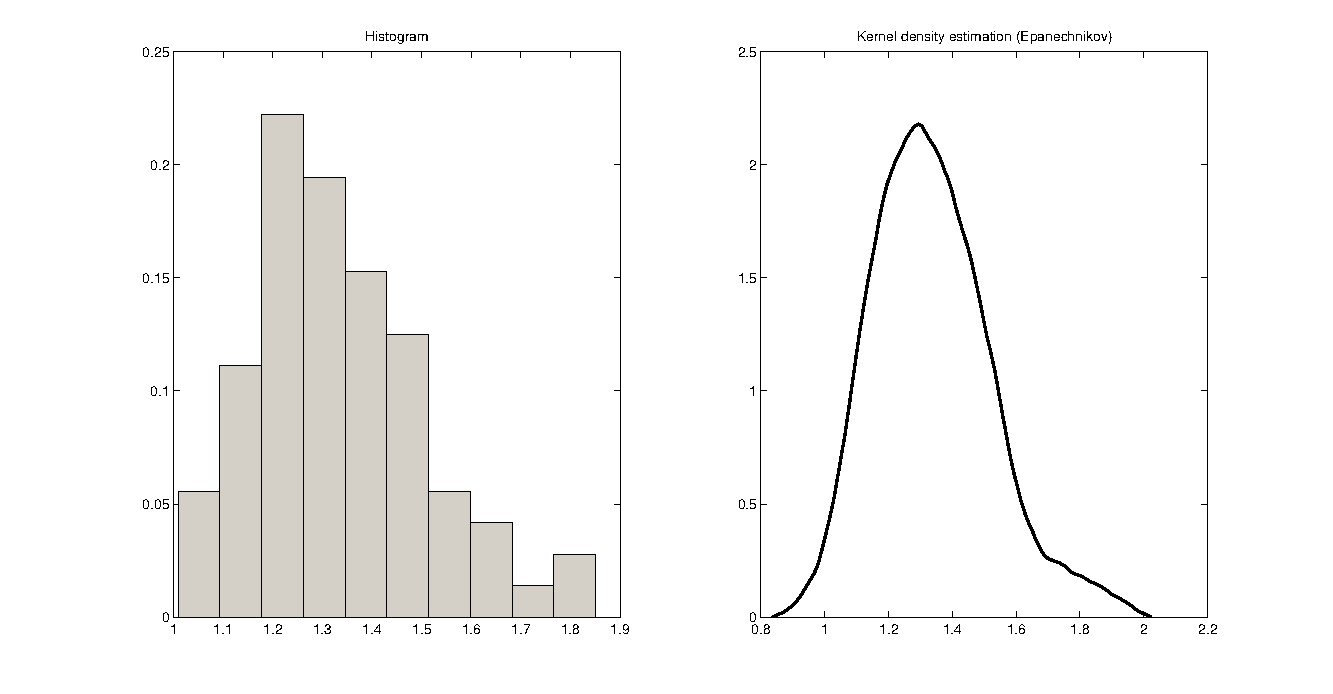}
	\caption{Example of histogram measurement on a generic $ith$ observation}
	\label{hist}
\end{figure}

\noindent Multiple valued data or Multiple Values Variables (MVV) are included in the framework of symbolic data \citep*{billard, bock}. The data descriptions of the units are called \emph{symbolic} when they are more complex than the standard ones due to the fact that they contain internal variation and are structured. Symbolic data need more complex data tables called \emph{symbolic data tables} because a cell of such data table does not necessarily contain as usual, a single quantitative or categorical values. The symbolic variables are usually represented as weight (probability) distributions or interval values.\\
Let $X$ be a continuous variable defined on a finite support $\mathcal{S} = \left[ \underline{x}, \overline{x}    \right]$, where $\underline{x}$ and $\overline{x}$ are the minimum and maximum values of the domain of $X$.
An histogram of $X$ is the representation of the Empirical Probability Distribution Function (EPDF), described by a set of pairs $\left(I_h, \pi_h \right)$, $h = 1,\ldots, H$, where $H$ is the number of contiguous intervals (bins) $\left\{ I_1, \ldots, I_h, \ldots, I_H \right\}$, where $I_h=[ \underline{x}_h;\overline{x}_h [$, in which the support $\mathcal{S}$ is partitioned and $\pi_h$ is the frequency associated with each interval. Figure \ref{hist} shows an example of such a situation, visualizing both the histogram and the kernel density estimation of the distribution.

\begin{figure}[h!]
	\centering
		\includegraphics[width=0.8\textwidth]{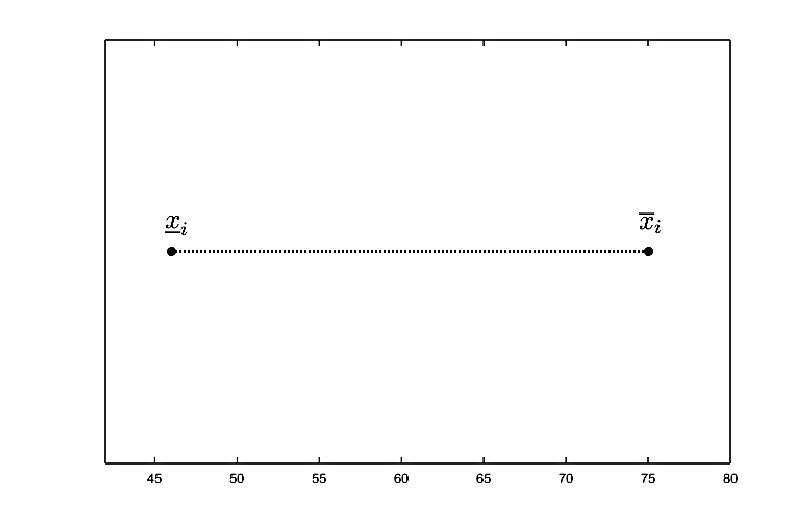}
	\caption{Example of interval measurement on a generic $ith$ observation}
	\label{int}
\end{figure}

\noindent A generic interval variable $X$ is a correspondence between a set $n$ of units and a set of closed intervals $\left[\underline{x}_{i}, \textrm{ }\overline{x}_{i}\right]$, with $i=1,\ldots,n$, $\underline{x}_{i} \leq \overline{x}_{i}$ and $\underline{x}_{i}, \overline{x}_{i} \in \Re$. An example characterized by such kind of interval can be represented as in Figure \ref{int}.

In dealing with multivalued data, in the literature tree-based methods are used with interval data as predictors by \cite*{mballo2}, and by \cite*{limam}. A preliminary pre-processing of interval data is mandatory to build the tree-based structure. This pre-processing consists either in considering the lower bound of each interval or the upper bound of each interval. Then a normal tree-growing procedure is done by taking as impurity measure the Kolmogorov-Smirnov measure. As alternative pre-processing of interval data, the mean value of each interval can be considered. Authors does not consider the possibility to have histogram data.

\section{A new partitioning definition}

\noindent Tree-growing method depends on the nature of both the response variable and the predictors. Response variable governs the choice of the impurity measure which is the heterogeneity index or entropy measure for a categorical response in classification trees. Predictors govern the way the partitioning variables are defined \citep*{harvesting, tutore}. In binary trees, these are known as splitting variables which are dichotomous variables. When dealing with standard data, the number of splitting variables to be generated by each predictor depends on the nature of the predictor itself (i.e. numerical, ordinal or nominal). A numerical or ordinal predictor with $m$ distinct values provides $m-1$ candidate splitting variables of the type either $X \leq c$ or $X > c$ where $c$ is the cutpoint that can be any of the distinct values until the $m-1$-th ordered values. For a nominal predictor with $m$ distinct categories there are $2^{m-1}-1$ candidate splitting variables to be generated as all possible combinations of the predictor's categories into two groups.\\
\indent When dealing with a predictor matrix including not only point value data but also multivalued data or multiple values variables, it is necessary for these variables to determine how the partitioning variables can be generated.\\
\noindent Let $Y$ be the $n$-dimensional vector of the response variable describing the a-priori class assignment on the sample of $n$ objects. Let $\mathbf{\Gamma}$ be the $n\times Q$ matrix of $Q$ Multiple Valued Variables observed for the sample of $n$ objects. Let $\mathbf{Z}$ be the $n\times K$ matrix of $K$ predictors of numerical or categorical type. Let $\bold{X} = \left[\mathbf{\Gamma} \textrm{ }  \bf{Z}\right]$ be the predictors matrix of dimension $n\times P$, with $P=K+Q$. Let $X^p$ be the generic $p$-th predictor, with $p=1,\ldots,P$.\\
\newline
\noindent \textit{Partitioning by histogram descriptors}\\
\newline
\noindent Suppose that $X^p$ is represented by a MVV of the type \textit{histogram data}. For the predictor $X^p$ let $F_{X^p\_i}(u)$ be the ECDF (Empirical Cumulative Distribution Function) associated to the $i$-th object with a size $H_i$, and let $F_{X^p\_j}(u)$ be the ECDF associated to the $j$-th object with a size $H_j$. Main idea is to compare the two distributions in order to understand if the two objects can be considered to belong to the same group or not through a non parametric test. On this purpose, we assume the $i$-th object and its distribution as reference one, and we consider the Wilcoxon test statistic such to compare, step by step, any $j$th object with the $i$-th object:

\begin{equation}
\label{W}
W = \sum\limits_{h = 1}^{H_i} {r\left( {X^p\_{ij} } \right)}, 
\end{equation}

\noindent where $X^p\_{ij}$ takes into account the ordered and combined sample formed by merging both the two distributions of $X^p\_i$ and $X^p\_j$ and $r\left( X^p\_{ij}  \right)$ is the rank of $X^p\_{ij}$. It is well known that for inferential purposes, it is convenient the studentized version of the Wilcoxon test statistic as defined by the T-statistic

\begin{equation}
\label{T}
T  = \frac{{W - {{H_i\left( {H_j + H_i - 1} \right)} \mathord{\left/
 {\vphantom {{n\left( {H_j + H_i - 1} \right)} {2 - {1 \mathord{\left/
 {\vphantom {1 2}} \right.
 \kern-\nulldelimiterspace} 2}}}} \right.
 \kern-\nulldelimiterspace} {2 - {1 \mathord{\left/
 {\vphantom {1 2}} \right.
 \kern-\nulldelimiterspace} 2}}}}}{{\sqrt {{{mn\left( {H_j + H_i + 1} \right)} \mathord{\left/
 {\vphantom {{mn\left( {H_j + H_i + 1} \right)} {12}}} \right.
 \kern-\nulldelimiterspace} {12}}} }} \stackrel{d}{\rightarrow} N\left( {0,1} \right),
\end{equation}

\noindent which converges in distribution to the standard normal distribution.\\
The partitioning of objects at a given node is based on a set of ternary questions of the form:
\begin{enumerate}
\item Is $F_{X^p\_i}(u)<F_{X^p\_j}(u)$?,
\item Is $F_{X^p\_i}(u)>F_{X^p\_j}(u)$?,
\item Is $F_{X^p\_i}(u)=F_{X^p\_j}(u)$?,
\end{enumerate}
$\forall i\neq j$, $j=1,\ldots,n$.

\noindent The answer to these questions is given by a joint lecture of both T-statistics and the connected $p-value$. Consider we are using the $i$-th object as reference one, and we are deciding in which child node will fall down the $j$-th object. Indeed if $T_{j|i}<0$ and $p-value<\alpha$, then we are considering the first case and $j$-th object goes down in the left child node. On the other hand, if $T_{j|i}>0$ and $p-value<\alpha$, then we are considering the second case and $j$-th object goes down in the right child node. If $p-value>\alpha$ we are considering the third case, and $j$-th object goes down in the central child node.\\
We can conclude that, if there are $n$ distinct histograms the number of possible partitions to be generated is equal to $n$.\\
\noindent The splitting rule for histogram descriptors can be summarized as follow:

\[
\textrm{Splitting rule: }\left\{
\begin{array}{ll}
T_{j|i}<0 \cap \textrm{ P-value}<\alpha: & \textrm{ \emph{j}-th obj.} \rightarrow \textrm{ left child node;} \\ 	
T_{j|i}>0  \cap \textrm{ P-value}<\alpha: &\textrm{ \emph{j}-th obj.} \rightarrow \textrm{ right child node;}\\
\textrm{otherwise: } & \textrm{ \emph{j}-th obj.} \rightarrow \textrm{ central child node.}	
\end{array}
\right.
\]

\noindent \textit{Partitioning by interval descriptors}\\
\newline
\noindent Suppose now $X^p$ is the $p$-th predictor in the data matrix, and it is represented by MVV of the type \textit{interval data}. Let $\underline{x}^p_{i}$ and $\overline{x}^p_{i}$ be respectively the lower and the upper bound of the interval of the $i$-th object of the predictor $X^p$. Let $\underline{x}^p_{j}$ and $\overline{x}^p_{j}$ be respectively the lower and the upper bound of the interval of the $j$-th instance of the predictor $X^p$.\\
With respect to the $i$-th object, the following splitting rule is defined:

\[
\textrm{Splitting rule: }\left\{
\begin{array}{ll}
\textrm{ }\underline{x}^p_{j} < \underline{x}^p_{i}  \cap \overline{x}^p_{j} < \overline{x}^p_{i} & \textrm{ \emph{j}-th obj.} \rightarrow \textrm{ left child node;} \\ 	
\textrm{ }
\\
\textrm{ }\underline{x}^p_{j} > \underline{x}^p_{i}  \cap \overline{x}^p_{j} > \overline{x}^p_{i} &\textrm{ \emph{j}-th obj.} \rightarrow \textrm{ right child node;}\\
\textrm{ }
\\
\begin{array}{c}
\underline{x}^p_{j} \geq \underline{x}^p_{i} \cap \overline{x}^p_{j} \leq \overline{x}^p_{i} \\
\cup \\
\underline{x}^p_{j} \leq \overline{x}^p_{i} \cap \overline{x}^p_{j} \geq \overline{x}^p_{i}
\end{array}
 & \textrm{ \emph{j}-th obj.} \rightarrow \textrm{ central child node.}	
\end{array}
\right.
\]

\noindent In the first case the $j$-th object goes down in the left child node, in the second case $j$-th object goes down in the right child node, in the third case $j$-th object goes down in the central child node. As in the case of histogram data, we can conclude that, if there are $n$ distinct intervals the number of possible partitions to be generated is equal to $n-v+1$, with $v=$ number of intervals included in the third case above defined. Note that in the case of interval descriptors we need not a formal hypothesis testing to declare which of the three cases governs the splitting rule.

\subsection{D-CLASS TREE algorithm}

\noindent A distinction can be done between standard data and multivalued data in the way the set of partitioning variables is generated.  
\noindent In presence of standard data, partitioning variables are generated as usual yielding to a binary split of the objects. In defining a set of ternary questions, it is clear that one of the properties generally accepted in the definition of a tree-based structure is removed by following this approach. Indeed, the \emph{shape property} is no more present when a tree-based methods involve predictors of the type MVV. According our opinion, this is necessary to preserve the goodness of tree-interpretability.\\
Once that the set of candidate partitioning variables has been generated, the best partition into either two or three subgroups at a given node can be selected maximizing the decrease in impurity $\Delta i(t,s)$:

\begin{equation}
	\max_{s} \left[\Delta i(t,s)\right] = \max_{s}\left[ i(t)-\sum_{k \in l,c,r}i_s(t_k)p_k\right],
\end{equation}

\noindent where the subscript $k \in l,c,r$ indicates respectively the left, central and right descendant node, and $p_k$ is referred to the proportion of objects that from the $t$-th node falls down in the $t_k$-th child node. Furthermore, $i(t)$ denotes the impurity measure of node $t$ and $i_s(t_k)$ is the impurity of child node $t_k$ obtained by the split $s$.\\
\noindent It is straightforward that partitions generated by \emph{classical} predictors generate two children nodes as well as partitions generated by MVV generate three children nodes. Tree-growing procedure ends when a stopping rule occurs. In general, stopping rules involve a maximum depth of the tree, a bound in the decrease in impurity or a bound in the sample size within node.\\
The innovative contribution of our algorithm refers to tree-growing procedure, specifically it refers to a new way to define the splitting variables, namely the set of each cutting points generated by a givn predictor. With respect to explorative purposes, it means that the interpretability of partitions takes in account a more rigorous information when MVV predictors generate splits. About decisional purposes, none is changed with respect to classical approaches. Indeed both division of the total sample in learning sample and test sample and cross-validation procedures are possible. Our approach allows to such a classifier to preserve the conditions to be used with ensemble methods such as Bagging, Boosting, Random Forests, etc. \citep*{bagging, forest, boosting}.

\section{A real world case}

\noindent D-CLASS TREE has been performed on a database of the Department of Dermatology of the Second University of Naples. The database consists of $220$ skin lesion dermoscopic images, for which a histological diagnosis is available, with a resolution of $768 \times 512$ pixels, divided into two classes: $86$ images are relative to malignant melanoma and $134$ of these lesions are classified as benign lesions. The skin lesion dermoscopic images are acquired using a charge-coupled devise camera connected to an epiluminescence microscopy. Data analyses were performed with our own routines written in MatLab language on a Computer Intel Core i5-3317U 1.70 GHz and 4GB of RAM.

\subsection{Skin lesions data set}

\noindent The data set consists in $34$ variables or \emph{descriptors} (including $11$ point values, $6$ intervals data and $17$ histograms data), plus a binary response variable. The multi-valued data describing the dermoscopic image database is structured as a matrix $D = \left\{ d_{i,p} \right\}$, where the rows represent the statistical units, i.e. the images, and the columns represent the multi-valued descriptors. Each matrix cell $d_{i,p}$ indicates the set of values attained by the $i$-th image for the $p$-th descriptor, that can be a scalar real value, an interval value, or a set of histogram values. A gray scale digital image is a two-dimensional discrete function $f\left(x, y\right)$, where $x$ and $y$ are spatial discrete coordinates, and the amplitude of $f$ at any pair of coordinates $\left(x, y\right)$ is the gray level of the image at that point, usually ranging in the interval $\left[0,255\right]$. Each element of a digital image, having a particular location and value, is usually referred to as pixel (picture element). Color (RGB) digital images are usually represented by three gray
scale images, one for each of the three Red, Green, and Blue primary color components; therefore, each pixel is a vector of three RGB scalar values. Any dermoscopic image retains information concerning physical characteristics of the skin lesion, such as colors and shape. In order to extract descriptors that represent such information, scalar or vector data contained
in each pixel should be properly aggregated and combined, and suitable measures should be chosen on each ensemble. For example, the average color of the skin lesion can be computed as the mean color of image pixels belonging to the lesion area. Therefore, the first digital processing step for dermoscopic images consists in segmenting the images in order to separate the skin area
and the lesion area. The segmentation has been achieved by the Otsu algorithm, that computes the optimum threshold separating the two classes of pixels (skin and lesion) so that their intra-class variance is minimal \citep*{otsu}. Following the ABCD-rule of dermoscopy \citep*{nachbar}, descriptors chosen for characterizing different lesion classes consist of quantitative measures of asymmetry, border, and color information extracted by dermoscopic images. More details about ABCD-rule can be found, in example, in \cite*{abcd, maglogiannis, celebi}. Table \ref{descr} summarizes the descriptors of the skin lesions data set

\begin{table}[!h!]
\tiny
\centering
\caption{List of descriptors. Point = real scalar; Int = interval data; Hist = histogram data}
\begin{tabular}{|l|l|l|l}
\cline{1-3}
\multicolumn{1}{|c|}{\textbf{Descriptor}} & \multicolumn{1}{c|}{\textbf{Nature}} & \multicolumn{1}{c|}{\textbf{Legend}} &  \\ 
\cline{1-3}
Area & \multicolumn{1}{c|}{Point} & Lesion area &  \\ 
\cline{1-3}
Perimeter & \multicolumn{1}{c|}{Point} & Lesion perimeter &  \\ 
\cline{1-3}
EquivDiameter & \multicolumn{1}{c|}{Point} & Equivalent diameter &  \\ 
\cline{1-3}
Eccentricity & \multicolumn{1}{c|}{Point} & Eccentricity &  \\ 
\cline{1-3}
MinMaxAxisLength & \multicolumn{1}{c|}{Int} & Major and minor interval axis &  \\ 
\cline{1-3}
AsymmXY & \multicolumn{1}{c|}{Int} & Shape asymmetry &  \\ 
\cline{1-3}
AsymmCelebiA1 & \multicolumn{1}{c|}{Point} & Alternative shape asymmetry &  \\ 
\cline{1-3}
AsymmCelebiA2 & \multicolumn{1}{c|}{Point} & Alternative shape asymmetry &  \\ 
\cline{1-3}
AsymmXYRed & \multicolumn{1}{c|}{Int} & Red asymmetry &  \\ 
\cline{1-3}
AsymmXYGreen & \multicolumn{1}{c|}{Int} & Green asymmetry &  \\ 
\cline{1-3}
AsymmXYBlue & \multicolumn{1}{c|}{Int} & Blue asymmetry &  \\ 
\cline{1-3}
MagloZoneExt & \multicolumn{1}{c|}{Hist} & Pixel average intensity within external section &  \\ 
\cline{1-3}
MagloZoneMid & \multicolumn{1}{c|}{Hist} & Pixel average intensity within middle section &  \\ 
\cline{1-3}
MagloZoneInt & \multicolumn{1}{c|}{Hist} & Pixel average intensity within internal section &  \\ 
\cline{1-3}
Degradation1 & \multicolumn{1}{c|}{Hist} & Color degradation &  \\ 
\cline{1-3}
Degradation2 & \multicolumn{1}{c|}{Hist} & Alternative color degradation &  \\ 
\cline{1-3}
Degradation3 & \multicolumn{1}{c|}{Hist} & Alternative color degradation &  \\ 
\cline{1-3}
Degradation4 & \multicolumn{1}{c|}{Hist} & Alternative color degradation &  \\ 
\cline{1-3}
Degradation5 & \multicolumn{1}{c|}{Hist} & Alternative color degradation &  \\ 
\cline{1-3}
Degradation6 & \multicolumn{1}{c|}{Hist} & Alternative color degradation &  \\ 
\cline{1-3}
Compactness & \multicolumn{1}{c|}{Point} & Compactness index &  \\ 
\cline{1-3}
MinMaxBorderDist & \multicolumn{1}{c|}{Int} & Minimum and maximum distance border bari-center &  \\ 
\cline{1-3}
RappBorderDist & \multicolumn{1}{c|}{Point} & Ratio between minimum and maximum distance border  bari-center &  \\ 
\cline{1-3}
VectorDistNorm & \multicolumn{1}{c|}{Hist} & Distances border bari-center &  \\ 
\cline{1-3}
CVBorderDist & \multicolumn{1}{c|}{Point} & Coefficient of variation distance border bari-center &  \\ 
\cline{1-3}
IntBord1 & \multicolumn{1}{c|}{Hist} & Border interruption &  \\ 
\cline{1-3}
IntBord2 & \multicolumn{1}{c|}{Hist} & Alternative border interruption &  \\ 
\cline{1-3}
IntBord3 & \multicolumn{1}{c|}{Hist} & Alternative border interruption &  \\ 
\cline{1-3}
IntBord4 & \multicolumn{1}{c|}{Hist} & Alternative border interruption &  \\ 
\cline{1-3}
IntBord5 & \multicolumn{1}{c|}{Hist} & Alternative border interruption &  \\ 
\cline{1-3}
IntBord6 & \multicolumn{1}{c|}{Hist} & Alternative border interruption &  \\ 
\cline{1-3}
IntBord7 & \multicolumn{1}{c|}{Hist} & Alternative border interruption &  \\ 
\cline{1-3}
Smoothness & \multicolumn{1}{c|}{Point} & Smoothness index &  \\ 
\cline{1-3}
Solidity & \multicolumn{1}{c|}{Point} & Solidity index &  \\ 
\cline{1-3}
\end{tabular}
	\label{descr}
\end{table}

\subsection{Data analysis}

\begin{figure}[h!]
	\centering
		\includegraphics[width=1\textwidth]{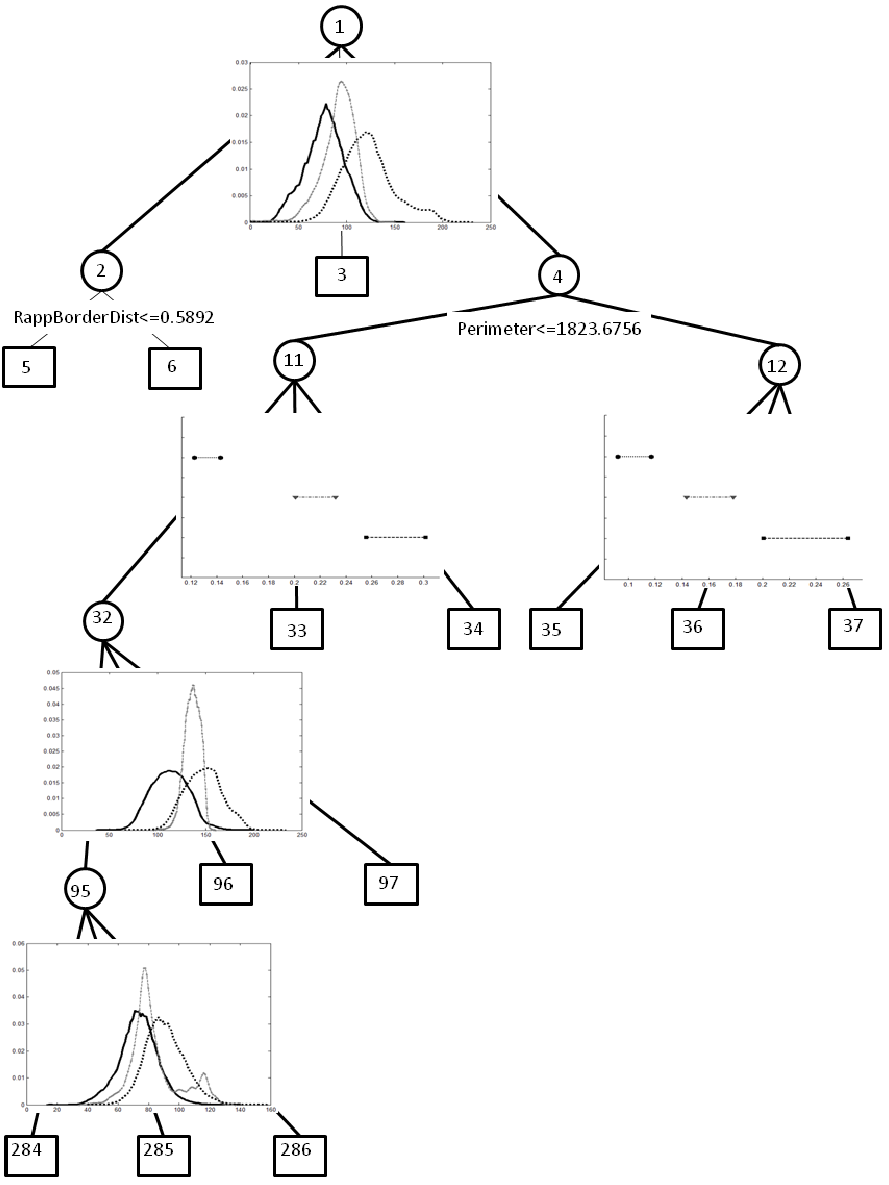}
	\caption{D-CLASS TREE}
	\label{ternary}
\end{figure}

\noindent Figure \ref{ternary} shows the D-CLASS TREE for the data previously described. For the enumeration of the nodes we used the following rule:

\begin{equation}
\#t_f = 3\left( {\# f - 1} \right) + \left\{ \begin{array}{l}
 2 \\ 
 3 \\ 
 4 \\ 
 \end{array} \right.
\end{equation}
where $\# t_f$ is the number of the node to be computed and $\# f$ indicates the number of its father node.

\noindent The figure emphasizes the way the splits are generated. For ternary splits generated by a histogram variable, a plot showing the Kernel density function estimate of the typical distributions is plotted within the graph. The central density function (in light grey) refers to the distributions going down in the central child node. The left and the right denstity functions (respectively bold-black and dot-black) refer to the distributions going down respectively in the left and right children nodes. If the splitting variable is generated by interval data, a plot showing these intervals is put in the graph. The central interval refers to images going down in the central child node, as well as upper-left and lower-right intervals refer to images going down respectively to the left and right children nodes. If the splitting variable is generated by point variables, the split is binary and in the figure is indicated the cutting point.
The error rate (misclassification error) at root node is equal to $0.3909$ as well as error rate of the tree is equal to $0.1909$.\\
\noindent Table \ref{treedes} shows the DCTree in tabular format. First four columns indicate respectively the node number, the node size, the children nodes generated by the actual node and the father of the actual node. Column, named splitting predictor, indicates predictor generates the split. In parenthesis the nature of the predictor is indicated ((H) histogram, (I) interval, (P) point). The column named splitting characteristics describe the split. If the splitting predictor is a histogram data, then some descriptive information about the distribution of the reference image is reported in the column (precisely Min, Max, Mean, Standard Deviation, Skewness, Kurtosis).  If the splitting predictor is a interval data, then upper and lower bounds of the reference interval are respectively reported in brackets in the column. If the splitting predictor is a point variable, then the cutting point is reported in the column.

\begin{table}[!h!]
\small
\centering
\caption{D-CLASS TREE results}
	\vspace{0.2cm}
\scalebox{0.85}{
\begin{tabular}{|l|l|l|l|lll|l|l|}
\hline
\multicolumn{1}{|c|}{Node} &  \multirow{2}{*}{Size} & \multirow{2}{*}{Children} & \multirow{2}{*}{Father} & \multicolumn{1}{c|}{Splitting} & \multicolumn{2}{c|}{Splitting} & \multirow{2}{*}{Rt} & \multirow{2}{*}{Class} \\ 
\multicolumn{1}{|c|}{number} & \multicolumn{1}{c|}{} & \multicolumn{1}{c|}{} & \multicolumn{1}{c|}{} & \multicolumn{1}{c|}{predictor} & \multicolumn{2}{c|}{characteristics} & \multicolumn{1}{c|}{} & \multicolumn{1}{c|}{} \\ 
\hline
\multicolumn{1}{|c|}{} & \multicolumn{1}{c|}{} & \multicolumn{1}{c|}{} & \multicolumn{1}{c|}{} & \multicolumn{1}{l|}{} & \multicolumn{1}{r}{$0.33$ } & \multicolumn{1}{r|}{$145.66$} & \multicolumn{1}{r|}{} & \multicolumn{1}{c|}{} \\ 
\multicolumn{1}{|c|}{$1$} & \multicolumn{1}{c|}{$220$} & \multicolumn{1}{c|}{$2$ $3$ $4$ } & \multicolumn{1}{c|}{-} & \multicolumn{1}{l|}{MagloZoneExt (H)} & \multicolumn{1}{r}{$91.34$} & \multicolumn{1}{r|}{$17.74$} & \multicolumn{1}{r|}{$0.39$} & \multicolumn{1}{c|}{B} \\ 
\multicolumn{1}{|c|}{} & \multicolumn{1}{c|}{} & \multicolumn{1}{c|}{} & \multicolumn{1}{c|}{} & \multicolumn{1}{l|}{} & \multicolumn{1}{r}{$-1.07$ } & \multicolumn{1}{r|}{$5.51$} & \multicolumn{1}{r|}{} & \multicolumn{1}{c|}{} \\ 
\hline
\multicolumn{1}{|c|}{$2$} & \multicolumn{1}{c|}{$65$} & \multicolumn{1}{c|}{$5$ $6$} & \multicolumn{1}{c|}{$1$} & \multicolumn{1}{l|}{RappBorderDist (P)} & \multicolumn{2}{c|}{$0.589$} & \multicolumn{1}{r|}{$0.20$} & \multicolumn{1}{c|}{M} \\ 
\hline
\multicolumn{1}{|c|}{$3$} & \multicolumn{1}{c|}{$3$} & \multicolumn{1}{c|}{-} & \multicolumn{1}{c|}{$1$} & \multicolumn{1}{c|}{Terminal} & \multicolumn{2}{c|}{-} & \multicolumn{1}{r|}{$0.00$} & \multicolumn{1}{c|}{M} \\ 
\hline
\multicolumn{1}{|c|}{$4$} & \multicolumn{1}{c|}{$152$} & \multicolumn{1}{c|}{$11$ $12$} & \multicolumn{1}{c|}{$1$} & \multicolumn{1}{l|}{Perimeter (P)} & \multicolumn{2}{c|}{$1823.67$} & \multicolumn{1}{r|}{$0.21$} & \multicolumn{1}{c|}{B} \\ 
\hline
\multicolumn{1}{|c|}{$5$} & \multicolumn{1}{c|}{$46$} & \multicolumn{1}{c|}{-} & \multicolumn{1}{c|}{$2$} & \multicolumn{1}{c|}{Terminal} & \multicolumn{2}{c|}{-} & \multicolumn{1}{r|}{$0.06$} & \multicolumn{1}{c|}{M} \\ 
\hline
\multicolumn{1}{|c|}{$6$} & \multicolumn{1}{c|}{$19$} & \multicolumn{1}{c|}{-} & \multicolumn{1}{c|}{$2$} & \multicolumn{1}{c|}{Terminal} & \multicolumn{2}{c|}{-} & \multicolumn{1}{r|}{$0.37$} & \multicolumn{1}{c|}{B} \\ 
\hline
\multicolumn{1}{|c|}{$11$} & \multicolumn{1}{c|}{$95$} & \multicolumn{1}{c|}{$32$ $33$ $34$} & \multicolumn{1}{c|}{$4$} & \multicolumn{1}{l|}{AsymmXYRed (I)} & \multicolumn{1}{r}{$[0.20$} & \multicolumn{1}{r|}{$0.23]$} & \multicolumn{1}{r|}{$0.06$} & \multicolumn{1}{c|}{B} \\ 
\hline
\multicolumn{1}{|c|}{$12$} & \multicolumn{1}{c|}{$57$} & \multicolumn{1}{c|}{$35$ $36$ $37$} & \multicolumn{1}{c|}{$4$} & \multicolumn{1}{l|}{AsymmXY (I)} & \multicolumn{1}{r}{$[0.14$} & \multicolumn{1}{r|}{$0.18]$} & \multicolumn{1}{r|}{$0.44$} & \multicolumn{1}{c|}{B} \\ 
\hline
\multicolumn{1}{|c|}{} & \multicolumn{1}{c|}{} & \multicolumn{1}{c|}{} & \multicolumn{1}{c|}{} & \multicolumn{1}{l|}{} & \multicolumn{1}{r}{$98.67$} & \multicolumn{1}{r|}{$157.33$} & \multicolumn{1}{r|}{} & \multicolumn{1}{c|}{} \\ 
\multicolumn{1}{|c|}{$32$} & \multicolumn{1}{c|}{$86$} & \multicolumn{1}{c|}{$95$ $96$ $97$} & \multicolumn{1}{c|}{$11$} & \multicolumn{1}{l|}{MagloZoneExt (H)} & \multicolumn{1}{r}{$136.04$} & \multicolumn{1}{r|}{$8.09$} & \multicolumn{1}{r|}{$0.03$} & \multicolumn{1}{c|}{B} \\ 
\multicolumn{1}{|c|}{} & \multicolumn{1}{c|}{} & \multicolumn{1}{c|}{} & \multicolumn{1}{c|}{} & \multicolumn{1}{l|}{} & \multicolumn{1}{r}{$-0.44$} & \multicolumn{1}{r|}{$3.21$} & \multicolumn{1}{r|}{} & \multicolumn{1}{c|}{} \\ 
\hline
\multicolumn{1}{|c|}{$33$} & \multicolumn{1}{c|}{$6$} & \multicolumn{1}{c|}{-} & \multicolumn{1}{c|}{$11$} & \multicolumn{1}{c|}{Terminal} & \multicolumn{2}{c|}{-} & \multicolumn{1}{r|}{$0.33$} & \multicolumn{1}{c|}{M} \\ 
\hline
\multicolumn{1}{|c|}{$34$} & \multicolumn{1}{c|}{$3$} & \multicolumn{1}{c|}{-} & \multicolumn{1}{c|}{$11$} & \multicolumn{1}{c|}{Terminal} & \multicolumn{2}{c|}{-} & \multicolumn{1}{r|}{$0.00$} & \multicolumn{1}{c|}{B} \\ 
\hline
\multicolumn{1}{|c|}{$35$} & \multicolumn{1}{c|}{$23$} & \multicolumn{1}{c|}{-} & \multicolumn{1}{c|}{$12$} & \multicolumn{1}{c|}{Terminal} & \multicolumn{2}{c|}{-} & \multicolumn{1}{r|}{$0.35$} & \multicolumn{1}{c|}{B} \\ 
\hline
\multicolumn{1}{|c|}{$36$} & \multicolumn{1}{c|}{$11$} & \multicolumn{1}{c|}{-} & \multicolumn{1}{c|}{$12$} & \multicolumn{1}{c|}{Terminal} & \multicolumn{2}{c|}{-} & \multicolumn{1}{r|}{$0.09$} & \multicolumn{1}{c|}{B} \\ 
\hline
\multicolumn{1}{|c|}{$37$} & \multicolumn{1}{c|}{$23$} & \multicolumn{1}{c|}{-} & \multicolumn{1}{c|}{$12$} & \multicolumn{1}{c|}{Terminal} & \multicolumn{2}{c|}{-} & \multicolumn{1}{r|}{$0.29$} & \multicolumn{1}{c|}{M} \\ 
\hline
\multicolumn{1}{|c|}{} & \multicolumn{1}{c|}{} & \multicolumn{1}{c|}{} & \multicolumn{1}{c|}{} & \multicolumn{1}{l|}{} & \multicolumn{1}{r}{$38.00$} & \multicolumn{1}{r|}{$133.33$} & \multicolumn{1}{r|}{} & \multicolumn{1}{c|}{} \\ 
\multicolumn{1}{|c|}{$95$} & \multicolumn{1}{c|}{$73$} & \multicolumn{1}{c|}{$284$ $285$ $286$} & \multicolumn{1}{c|}{$32$} & \multicolumn{1}{l|}{MagloZoneInt (H)} & \multicolumn{1}{r}{$83.47$} & \multicolumn{1}{r|}{$15.90$} & \multicolumn{1}{r|}{$0.03$} & \multicolumn{1}{c|}{B} \\ 
\multicolumn{1}{|c|}{} & \multicolumn{1}{c|}{} & \multicolumn{1}{c|}{} & \multicolumn{1}{c|}{} & \multicolumn{1}{l|}{} & \multicolumn{1}{r}{$0.87$} & \multicolumn{1}{r|}{$3.36$} & \multicolumn{1}{r|}{} & \multicolumn{1}{c|}{} \\ 
\hline
\multicolumn{1}{|c|}{$96$} & \multicolumn{1}{c|}{$1$} & \multicolumn{1}{c|}{-} & \multicolumn{1}{c|}{$32$} & \multicolumn{1}{c|}{Terminal} & \multicolumn{2}{c|}{-} & \multicolumn{1}{r|}{$0.00$} & \multicolumn{1}{c|}{M} \\ 
\hline
\multicolumn{1}{|c|}{$97$} & \multicolumn{1}{c|}{$12$} & \multicolumn{1}{c|}{-} & \multicolumn{1}{c|}{$32$} & \multicolumn{1}{c|}{Terminal} & \multicolumn{2}{c|}{-} & \multicolumn{1}{r|}{$0.00$} & \multicolumn{1}{c|}{B} \\ 
\hline
\multicolumn{1}{|c|}{$284$} & \multicolumn{1}{c|}{$29$} & \multicolumn{1}{c|}{-} & \multicolumn{1}{c|}{$95$} & \multicolumn{1}{c|}{Terminal} & \multicolumn{2}{c|}{-} & \multicolumn{1}{r|}{$0.03$} & \multicolumn{1}{c|}{B} \\ 
\hline
\multicolumn{1}{|c|}{$285$} & \multicolumn{1}{c|}{$1$} & \multicolumn{1}{c|}{-} & \multicolumn{1}{c|}{$95$} & \multicolumn{1}{c|}{Terminal} & \multicolumn{2}{c|}{-} & \multicolumn{1}{r|}{$0.00$} & \multicolumn{1}{c|}{M} \\ 
\hline
\multicolumn{1}{|c|}{$286$} & \multicolumn{1}{c|}{$43$} & \multicolumn{1}{c|}{-} & \multicolumn{1}{c|}{$95$} & \multicolumn{1}{c|}{Terminal} & \multicolumn{2}{c|}{-} & \multicolumn{1}{r|}{$0.00$} & \multicolumn{1}{c|}{B} \\ 
\hline
\end{tabular}
}
	\label{treedes}
\end{table}

\noindent The last two columns refer to the misclassification ratio within node (Rt) and the assigned class within node (B=Benignant, M=Malignant).

\subsection{A comparative study}

\begin{figure}[h!]
	\centering
		\includegraphics[width=1\textwidth]{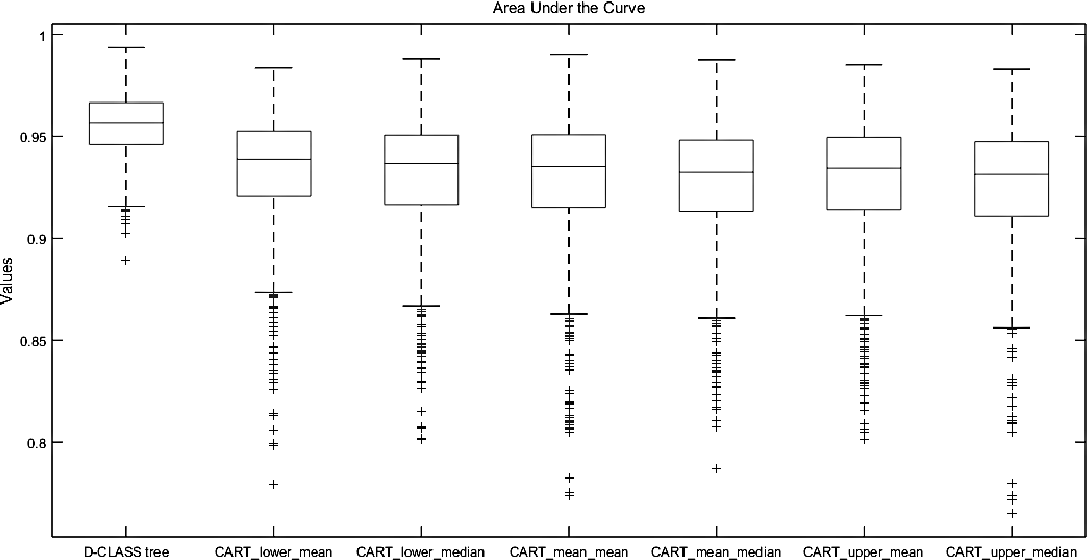}
	\caption{AUC distribution over 1,000 bootstrap replications}
	\label{AUC}
\end{figure}

\noindent This section is about a comparative study of the performance of the D-CLASS TREE with respect to the other tree-based classifiers dealing with MVV as predictors using the same real dataset. As in literature tree-based methods are used just for interval data, we adapted the algorithms described previously by computing for histogram data the mean value or the median value of each distribution. We built several classification trees: CART\_Lower\_Mean, CART\_Upper\_Mean, CART\_Lower\_Median, CART\_Upper\_Median, CART\_Mean\_Mean, CART\_Mean\_Median. The first word is referred to the processing of interval data, the second word is referred to the processing of histogram data. In example CART\_Lower\_Mean has as predictors the lower bound for interval data and the mean value for histogram descriptors. The experiment consisted in 1,000 bootstrap replication to test the stability of the tree-based structure and the performance of the algorithms in terms of Area Under The Curve, Brier Score and Misclassification ratio.\\
\noindent Specially in medical classification problems, the terms sensitivity and specificity are used to characterize a classification rule. The sensitivity is the probability of predicting a disease given that the true state is disease, as well as the specificity is the probability of predicting a non-disease given that the true state is non-disease.  Receiver Operating Characteristics (ROC) graphs are a useful technique for organizing classifiers and visualizing their performance. ROC graphs are two-dimensional graphs in which sensitivity rate is plotted on the vertical axis and $1-$specificity rate is plotted on the horizontal axis. While a ROC curve is a two-dimensional depiction of classifier performance, to compare different classifiers a way to proceed is to achieve a single scalar value representing the expected performance of each of them. A common method is to calculate the area under the ROC curve (AUC) \citep*{bradley}. The area under the ROC curve is sometimes called the \textit{c-statistic}. It can be shown that the area under the ROC curve is equivalent to the Mann-Whitney $U$ statistic for the median difference between the prediction scores in the two groups \citep*{statlearn}. Obviously, the higher is the AUC, the better is the performance of the classifier. Figure \ref{AUC} shows the performance of the D-CLASS Tree versus the other versions of the CART algorithm in terms of AUC. 

Box-plots concern the distribution of the area under the curve over 1,000 bootstrap replications. The boostrap samples were always different for each algorithm. The figure points out that the D-CLASS TREE returns better results in terms of AUC. In fact, it slightly outperforms their competitors. This result is emphasized by looking at both Tables \ref{kwallisAUC} and \ref{ttestAUC}. Table \ref{kwallisAUC} shows the one-way ANOVA performed by considering the different algorithms as factors. The analysis was carried out after transforming the data with the Box and Cox transformation \citep{boxcox}. After that transformation, a Kolmogorov-Smirnov test confirmed that transformed data can be assumed normally distributed (KS statistic = $0.0148$, P-value = $0.0915$). 

\begin{table}[h]
  \centering
  \caption{Area Under the Curve: ANOVA table after the Box-Cox transformation}
	  \label{kwallisAUC}%
	\vspace{0.2cm}
    \begin{tabular}{|c c c c c c|}
		\hline
		\multicolumn{1}{|c}{Source} & \multicolumn{1}{c}{SS} & \multicolumn{1}{c}{df} & \multicolumn{1}{c}{MS} & \multicolumn{1}{c}{F} & \multicolumn{1}{c|}{Sig.} \\
		\hline
    \multicolumn{1}{|c}{Groups} & \multicolumn{1}{l}{0.1145} & \multicolumn{1}{l}{6} & \multicolumn{1}{l}{0.0191} & \multicolumn{1}{l}{175.2800} & \multicolumn{1}{l|}{0.000}\\
    \multicolumn{1}{|c}{Error} & \multicolumn{1}{l}{0.7616} & \multicolumn{1}{l}{6993} & \multicolumn{1}{l}{0.0001} & \multicolumn{1}{}{} & \multicolumn{1}{l|}{}\\ 
		 \multicolumn{1}{|c}{Total} & \multicolumn{1}{l}{0.8761} & \multicolumn{1}{l}{6999} & \multicolumn{1}{l}{} & \multicolumn{1}{}{} & \multicolumn{1}{l|}{}\\ 
		\hline
		 \end{tabular}%
\end{table}%
		
Table \ref{ttestAUC} shows the multiple comparison tests by adopting the Bonferroni correction. Each cell contains the difference between the mean values and its associated p-value for each comparison. It is worth noting that always the Dynamic Classification Tree outperforms its competitors. It is worth highlighting that, except for few cases, other approaches do not contribute to explain the difference among the mean values as emphasized by the previous ANOVA analysis.

\begin{table}[h]
  \centering
  \caption{Area under the curve: multiple comparison test}
		\vspace{0.2cm}
    \begin{tabular}{|r|r|r|r|r|r|r|}
\cline{1-7}    \multicolumn{1}{|r|}{} & \multicolumn{1}{c|}{CART} & \multicolumn{1}{c|}{CART} & \multicolumn{1}{c|}{CART} & \multicolumn{1}{c|}{CART} & \multicolumn{1}{c|}{CART} & \multicolumn{1}{c|}{CART} \\
    \multicolumn{1}{|r|}{} & \multicolumn{1}{c|}{lower} & \multicolumn{1}{c|}{lower} & \multicolumn{1}{c|}{mean} & \multicolumn{1}{c|}{mean} & \multicolumn{1}{c|}{upper} & \multicolumn{1}{c|}{upper} \\
    \multicolumn{1}{|r|}{} & \multicolumn{1}{c|}{mean} & \multicolumn{1}{c|}{median} & \multicolumn{1}{c|}{mean} & \multicolumn{1}{c|}{median} & \multicolumn{1}{c|}{mean} & \multicolumn{1}{c|}{median} \\
		\hline
    \multicolumn{1}{|l|}{DCLASS\_tree} & 0.0098 & 0.0109 & 0.0110 & 0.0121 & 0.0115 & 0.0126 \\
    \multicolumn{1}{|c|}{Sig.} & \textbf{0.0000} & \textbf{0.0000} & \textbf{0.0000} & \textbf{0.0000} & \textbf{0.0000} & \textbf{0.0000} \\
    \hline
    \multicolumn{1}{|l|}{CART\_lower\_mean} & \multicolumn{1}{c|}{-} & 0.0011 & 0.0012 & 0.0023 & 0.0017 & 0.0028 \\
    \multicolumn{1}{|c|}{Sig.} & \multicolumn{1}{c|}{} & 0.3649 & 0.1636 & \textbf{0.0000} & \textbf{0.0058} & \textbf{0.0000} \\
    \hline
    \multicolumn{1}{|l|}{CART\_lower\_median} & \multicolumn{1}{c|}{-} & \multicolumn{1}{c|}{-} & 0.0000 & 0.0012  & 0.0000 & 0.0017 \\
    \multicolumn{1}{|c|}{Sig.} & \multicolumn{1}{c|}{} & \multicolumn{1}{c|}{} & 1.0000 & 0.2282 & 1.0000 & \textbf{0.0044} \\
    \hline
    \multicolumn{1}{|l|}{CART\_mean\_mean} & \multicolumn{1}{c|}{-} & \multicolumn{1}{c|}{-} & \multicolumn{1}{c|}{-} & 0.0011 & 0.0000 & 0.0016 \\
    \multicolumn{1}{|c|}{Sig.} & \multicolumn{1}{c|}{} & \multicolumn{1}{c|}{} & \multicolumn{1}{c|}{} & 0.4943 & 1.0000 & 0.0129 \\
    \hline
    \multicolumn{1}{|l|}{CART\_mean\_median} & \multicolumn{1}{c|}{-} & \multicolumn{1}{c|}{-} & \multicolumn{1}{c|}{-} & \multicolumn{1}{c|}{-} & 0.0000 & 0.0000 \\
    \multicolumn{1}{|c|}{Sig.} & \multicolumn{1}{c|}{} & \multicolumn{1}{c|}{} & \multicolumn{1}{c|}{} & \multicolumn{1}{c|}{} & 1.0000 & 1.0000 \\
    \hline
    \multicolumn{1}{|l|}{CART\_upper\_mean} & \multicolumn{1}{c|}{-} & \multicolumn{1}{c|}{-} & \multicolumn{1}{c|}{-} & \multicolumn{1}{c|}{-} & \multicolumn{1}{c|}{-} & 0.0011 \\
    \multicolumn{1}{|c|}{Sig.} &       &       &       &       &       & 0.2997 \\
    \hline
    \end{tabular}%
  \label{ttestAUC}%
\end{table}%

\noindent According to \cite*{ferri}, AUC measures are preferable in evaluating the performance of a classifier with small training datasets (as in our case). Nevertheless, the area under the ROC curve can be less useful for comparing methods when their ROC curves intercept, which can be the case, in example, for decision trees \citep*{zadrozny}. For this reason, we chose to use also the Brier score \citep*{brier}, also known as mean squared error \citep*{ferri}, as a method to check the goodness of the classifiers. Suppose we have just binary outcomes. Let $N$ be the number of instances, $f(i,j)$ the actual probablity of instance $i$ to be of class $j$, $j\in \{0,1\}$, and $o_i$ the reference to be equal to $0$ or $1$ depending on whether the a-priori classification is $0$ or $1$. Brier score is defined as

\begin{equation}
	BS=\frac{1}{n}\sum_{i=1}^n\left(f(i,j)-o_i\right)^2.
\end{equation}

\begin{figure}[h!]
	\centering
		\includegraphics[width=1\textwidth]{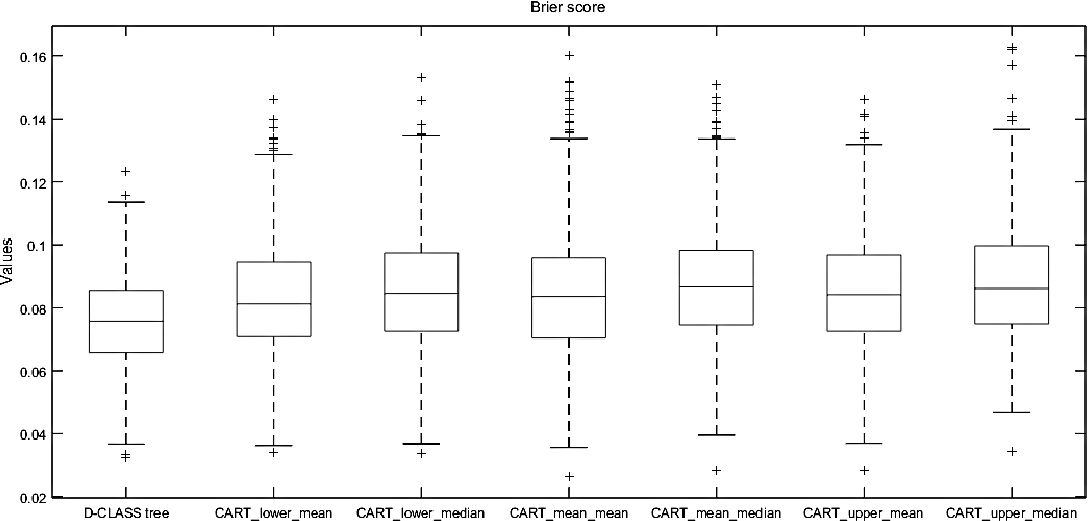}
	\caption{Brier score distribution over 1,000 bootstrap replications}
	\label{Brier}
\end{figure}

\noindent Figure \ref{Brier} shows the performance of the D-CLASS TREE versus the other versions of the CART algorithm in terms of Brier score. Unlike the previous analysis, the lower the score the better the performance. Box plot of the D-CLASS TREE highlights less variation around the median value of the Brier score over the 1,000 bootstrap replications. Moreover, the plot suggests that the D-CLASS TREE slightly outperforms its competitors.

This statement is confirmed by looking at both Tables \ref{krwallisbrier} and \ref{ttestbrier}. Also in this case we provided a one-way ANOVA and a multiple comparison test. We transformed the Brier scores with the Box and Cox transformation and checked the normal distribution with the Kolmogorov-Smirnov test (KS statistic = $0.0074$, P-value = $0.8347$). Table \ref{krwallisbrier} shows the ANOVA table.

\begin{table}[h]
  \centering
  \caption{Brier score: ANOVA table after the Box-Cox transformation}
	\vspace{0.2cm}
    \begin{tabular}{|c c c c c c|}
		\hline
		\multicolumn{1}{|c}{Source} & \multicolumn{1}{c}{SS} & \multicolumn{1}{c}{df} & \multicolumn{1}{c}{MS} & \multicolumn{1}{c}{F} & \multicolumn{1}{c|}{Sig.} \\
		\hline
    \multicolumn{1}{|c}{Groups} & \multicolumn{1}{l}{0.9606} & \multicolumn{1}{l}{6} & \multicolumn{1}{l}{0.1601} & \multicolumn{1}{l}{46.6000} & \multicolumn{1}{l|}{0.0000}\\
    \multicolumn{1}{|c}{Error} & \multicolumn{1}{l}{24.0217} & \multicolumn{1}{l}{6993} & \multicolumn{1}{l}{0.0034} & \multicolumn{1}{}{} & \multicolumn{1}{l|}{}\\ 
		 \multicolumn{1}{|c}{Total} & \multicolumn{1}{l}{24.9823} & \multicolumn{1}{l}{6999} & \multicolumn{1}{l}{} & \multicolumn{1}{l}{} & \multicolumn{1}{l|}{}\\ 
		\hline
		 \end{tabular}%
  \label{krwallisbrier}%
\end{table}%

Table \ref{ttestbrier} shows the multiple comparison tests considering the Bonferroni correction. The table highlights that always the Dynamic Classification Tree outperforms its competitors. Always the sign of the difference is negative and the p-value is highly significant. As in the analysis of the AUC, there are not classical algorithms systematically outperforming the others.

\begin{table}[h]
  \centering
  \caption{Brier score: multiple comparison test}
		\vspace{0.2cm}
    \begin{tabular}{|r|r|r|r|r|r|r|}
\cline{1-7}    \multicolumn{1}{|r|}{} & \multicolumn{1}{c|}{CART} & \multicolumn{1}{c|}{CART} & \multicolumn{1}{c|}{CART} & \multicolumn{1}{c|}{CART} & \multicolumn{1}{c|}{CART} & \multicolumn{1}{c|}{CART} \\
    \multicolumn{1}{|r|}{} & \multicolumn{1}{c|}{lower} & \multicolumn{1}{c|}{lower} & \multicolumn{1}{c|}{mean} & \multicolumn{1}{c|}{mean} & \multicolumn{1}{c|}{upper} & \multicolumn{1}{c|}{upper} \\
    \multicolumn{1}{|r|}{} & \multicolumn{1}{c|}{mean} & \multicolumn{1}{c|}{median} & \multicolumn{1}{c|}{mean} & \multicolumn{1}{c|}{median} & \multicolumn{1}{c|}{mean} & \multicolumn{1}{c|}{median} \\
		\hline
    \multicolumn{1}{|l|}{DCLASS\_tree} & -0.0231 & -0.0312 & -0.0270 & -0.0362 & -0.0298 & -0.0375 \\
    \multicolumn{1}{|c|}{Sig.} & \textbf{0.0000} & \textbf{0.0000} & \textbf{0.0000} & \textbf{0.0000} & \textbf{0.0000} & \textbf{0.0000} \\
    \hline
    \multicolumn{1}{|l|}{CART\_lower\_mean} & \multicolumn{1}{c|}{-} & -0.0080 & -0.0039 & -0.0131 & -0.0066 & -0.0143 \\
    \multicolumn{1}{|c|}{Sig.} & \multicolumn{1}{c|}{} & 0.0351 & 0.7624 & \textbf{0.0000} & 0.1496 & \textbf{0.0000} \\
    \hline
    \multicolumn{1}{|l|}{CART\_lower\_median} & \multicolumn{1}{c|}{-} & \multicolumn{1}{c|}{-} & 0.0042 & -0.0050 & 0.0014 & -0.0063 \\
    \multicolumn{1}{|c|}{Sig.} & \multicolumn{1}{c|}{} & \multicolumn{1}{c|}{} & 0.6840 & 0.4650 & 0.9982 & 0.2013 \\
    \hline
    \multicolumn{1}{|l|}{CART\_mean\_mean} & \multicolumn{1}{c|}{-} & \multicolumn{1}{c|}{-} & \multicolumn{1}{c|}{-} & -0.0092 & -0.0028 & -0.0105 \\
    \multicolumn{1}{|c|}{Sig.} & \multicolumn{1}{c|}{} & \multicolumn{1}{c|}{} & \multicolumn{1}{c|}{} & \textbf{0.0079} & 0.9406 & \textbf{0.0013} \\
    \hline
    \multicolumn{1}{|l|}{CART\_mean\_median} & \multicolumn{1}{c|}{-} & \multicolumn{1}{c|}{-} & \multicolumn{1}{c|}{-} & \multicolumn{1}{c|}{-} & 0.0065 & -0.0012 \\
    \multicolumn{1}{|c|}{Sig.} & \multicolumn{1}{c|}{} & \multicolumn{1}{c|}{} & \multicolumn{1}{c|}{} & \multicolumn{1}{c|}{} & 0.1731 & 0.9992 \\
    \hline
    \multicolumn{1}{|l|}{CART\_upper\_mean} & \multicolumn{1}{c|}{-} & \multicolumn{1}{c|}{-} & \multicolumn{1}{c|}{-} & \multicolumn{1}{c|}{-} & \multicolumn{1}{c|}{-} & -0.0077 \\
    \multicolumn{1}{|c|}{Sig.} &       &       &       &       &       & 0.0521 \\
    \hline
    \end{tabular}%
  \label{ttestbrier}%
\end{table}%

\begin{figure}[h!]
	\centering
		\includegraphics[width=1\textwidth]{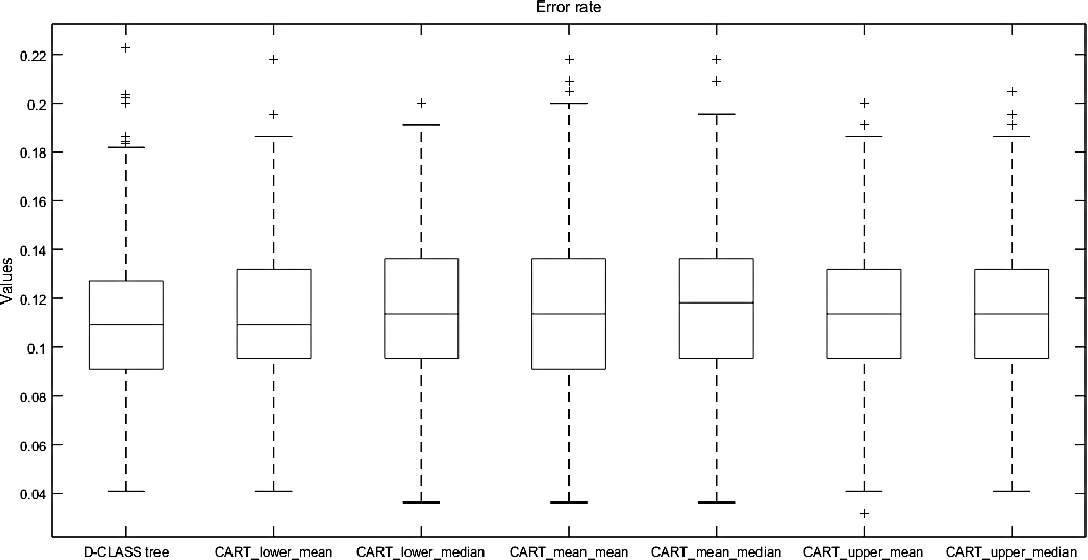}
		\caption{Error rates distribution over 1,000 bootstrap replications}
		\label{Rt}
\end{figure}

\noindent Figure \ref{Rt} shows the performance of the D-CLASS tree versus the other versions of the CART algorithm in terms of misclassification error over the 1,000 bootstrap replications. Looking at the box-plots it seems that none of the considered methods is outperformed by the competitors. 

In this case we performed a Kruskal-Wallis ANOVA \citep{kwanova}. We chose to use raw error rates because, even after the Box and Cox transformation, data were not normally distributed (Kolmogorov-Smirnov statistic = $0.0354$, p-value = $0.0000$). Table \ref{krwalliserror} leads us to reject the null hypothesis of equality between the means, but it is clear, by looking at Table \ref{rttest}, that there is not an algorithm systematically more performing than another.

\begin{table}[h]
  \centering
  \caption{Missclassification error: Kruskal-Wallis ANOVA table}
	\vspace{0.2cm}
    \begin{tabular}{|c c c c c c|}
		\hline
		\multicolumn{1}{|c}{Source} & \multicolumn{1}{c}{SS} & \multicolumn{1}{c}{df} & \multicolumn{1}{c}{MS} & \multicolumn{1}{c}{Chi-sq} & \multicolumn{1}{c|}{Sig.} \\
		\hline
    \multicolumn{1}{|c}{Groups} & \multicolumn{1}{l}{170.18e+06} & \multicolumn{1}{l}{6} & \multicolumn{1}{l}{283.63e+05} & \multicolumn{1}{l}{41.7400} & \multicolumn{1}{l|}{0.0000}\\
    \multicolumn{1}{|c}{Error} & \multicolumn{1}{l}{283.67e+08} & \multicolumn{1}{l}{6993} & \multicolumn{1}{l}{405.65e+04} & \multicolumn{1}{}{} & \multicolumn{1}{l|}{}\\ 
		 \multicolumn{1}{|c}{Total} & \multicolumn{1}{l}{283.37e+08} & \multicolumn{1}{l}{6999} & \multicolumn{1}{l}{} & \multicolumn{1}{l}{} & \multicolumn{1}{l|}{}\\ 
		\hline
		 \end{tabular}%
  \label{krwalliserror}%
\end{table}%

\begin{table}[h]
  \centering
  \caption{Missclassification error: multiple comparison test}
		\vspace{0.2cm}
    \begin{tabular}{|r|r|r|r|r|r|r|}
\cline{1-7}    \multicolumn{1}{|r|}{} & \multicolumn{1}{c|}{CART} & \multicolumn{1}{c|}{CART} & \multicolumn{1}{c|}{CART} & \multicolumn{1}{c|}{CART} & \multicolumn{1}{c|}{CART} & \multicolumn{1}{c|}{CART} \\
    \multicolumn{1}{|r|}{} & \multicolumn{1}{c|}{lower} & \multicolumn{1}{c|}{lower} & \multicolumn{1}{c|}{mean} & \multicolumn{1}{c|}{mean} & \multicolumn{1}{c|}{upper} & \multicolumn{1}{c|}{upper} \\
    \multicolumn{1}{|r|}{} & \multicolumn{1}{c|}{mean} & \multicolumn{1}{c|}{median} & \multicolumn{1}{c|}{mean} & \multicolumn{1}{c|}{median} & \multicolumn{1}{c|}{mean} & \multicolumn{1}{c|}{median} \\
		\hline
    \multicolumn{1}{|l|}{DCLASS\_tree} & -1.494 & -4.612 & -1.752 & -5.428 & -2.521 & -4.056 \\
    \multicolumn{1}{|c|}{Sig.} & 0.1353 & \textbf{0.0000} & 0.0816 & \textbf{0.0000} & 0.0117 & \textbf{0.0000} \\
    \hline
    \multicolumn{1}{|l|}{CART\_lower\_mean} & \multicolumn{1}{c|}{-} & -2932 & -0.327 & -3.772 & -0.947 & -2.372 \\
    \multicolumn{1}{|c|}{Sig.} & \multicolumn{1}{c|}{} & 0.0034 & 0.7434 & \textbf{0.0002} & 0.3435 & 0.0177 \\
    \hline
    \multicolumn{1}{|l|}{CART\_lower\_median} & \multicolumn{1}{c|}{-} & \multicolumn{1}{c|}{-} & 2.457 & -0.836 & 2.011 & 0.606 \\
    \multicolumn{1}{|c|}{Sig.} & \multicolumn{1}{c|}{} & \multicolumn{1}{c|}{} & 0.0140 & 0.4033 & 0.0143 & 0.5442 \\
    \hline
    \multicolumn{1}{|l|}{CART\_mean\_mean} & \multicolumn{1}{c|}{-} & \multicolumn{1}{c|}{-} & \multicolumn{1}{c|}{-} & -3.228 & -0.556 & -1.958 \\
    \multicolumn{1}{|c|}{Sig.} & \multicolumn{1}{c|}{} & \multicolumn{1}{c|}{} & \multicolumn{1}{c|}{} & 0.0012 & 0.5784 & 0.0502 \\
    \hline
    \multicolumn{1}{|l|}{CART\_mean\_median} & \multicolumn{1}{c|}{-} & \multicolumn{1}{c|}{-} & \multicolumn{1}{c|}{-} & \multicolumn{1}{c|}{-} & 2.829 & 1.439 \\
    \multicolumn{1}{|c|}{Sig.} & \multicolumn{1}{c|}{} & \multicolumn{1}{c|}{} & \multicolumn{1}{c|}{} & \multicolumn{1}{c|}{} & 0.0047 & 0.1499 \\
    \hline
    \multicolumn{1}{|l|}{CART\_upper\_mean} & \multicolumn{1}{c|}{-} & \multicolumn{1}{c|}{-} & \multicolumn{1}{c|}{-} & \multicolumn{1}{c|}{-} & \multicolumn{1}{c|}{-} & 1.441 \\
    \multicolumn{1}{|c|}{Sig.} &       &       &       &       &       & 0.1497 \\
    \hline
    \end{tabular}%
  \label{rttest}%
\end{table}%


\section{Concluding remarks}

In this paper, D-CLASS tree methodology for supervised classification has been introduced in order to deal with different types of data, both standard and multivalued data, namely point data, interval data, histogram data. A new partitioning criterion has been defined in the recursive partitioning yielding to a ternary tree structure. Main issue is the use of a Wilcoxon testing procedure to identify the partitioning variables on which basis to select the best one maximizing the decrease of impurity. 
D-CLASS TREE performance has been validated in a real world case study. A comparative study with respect to other tree-based classifiers dealing with standard and multivalued data demonstrate how D-CLASS TREE provides similar error rates estimates but it outperforms its competitors in terms of robustness of tree accuracy measured by both AUC and Brier score. This is an important result for predictive learning. Indeed we chose to use both AUC and Brier score approaches because they work with the (posterior) probability to belong to a given response class. Obviously, we read these probabilities (better, these indicator functions) by taking in account the distribution or the response variable in each terminal node. What we mean is that the Brier score (or the AUC), should be read together with the misclassification ratio because it measures the strength of the performance of a classifier, specially if the classification is made by a discrete classifier.

\section*{Acknowledgements}

Authors would like to thank Prof. A. Baroni of the Second University of Naples (Italy) for kindly providing us the Skin lesions data set.


\begin{thebibliography}{11}

\bibitem[Bashir et~al.(2014)Bashir, Qamar, and Khan]{bashir}
Bashir, S., Qamar, U., and Khan, F.~H.
\newblock Heterogeneous classifiers fusion for dynamic breast cancer diagnosis
  using weighted vote based ensemble.
\newblock \emph{Quality \& Quantity}, pages 1--16, 2014.

\bibitem[Billard and Diday(2003)]{billard}
Billard, L. and Diday, E.
\newblock From the statistics of data to the statistics of knowledge: symbolic
  data analysis.
\newblock \emph{Journal of the American Statistical Association}, 98\penalty0
  (462):\penalty0 470--487, 2003.

\bibitem[Bock and Diday(2012)]{bock}
Bock, H.-H. and Diday, E.
\newblock \emph{Analysis of symbolic data: exploratory methods for extracting
  statistical information from complex data}.
\newblock Springer Science \& Business Media, 2012.

\bibitem[Bono et~al.(1999)Bono, Tomatis, Bartoli, Tragni, Radaelli, Maurichi,
  and Marchesini]{abcd}
Bono, A., Tomatis, S., Bartoli, C., Tragni, G., Radaelli, G., Maurichi, A., and
  Marchesini, R.
\newblock The abcd system of melanoma detection.
\newblock \emph{Cancer}, 85\penalty0 (1):\penalty0 72--77, 1999.

\bibitem[Borgoni and Berrington(2013)]{borgoni}
Borgoni, R. and Berrington, A.
\newblock Evaluating a sequential tree-based procedure for multivariate
  imputation of complex missing data structures.
\newblock \emph{Quality \& Quantity}, 47\penalty0 (4):\penalty0 1991--2008,
  2013.


\bibitem[Box and Cox(1964)]{boxcox}
Box, G. E., and Cox, D. R.
\newblock An analysis of transformations. 
\newblock \emph{Journal of the Royal Statistical Society. Series B (Methodological)}, \penalty0 211--252, 1964.


\bibitem[Bradley(1997)]{bradley}
Bradley, A.~P.
\newblock The use of the area under the roc curve in the evaluation of machine
  learning algorithms.
\newblock \emph{Pattern recognition}, 30\penalty0 (7):\penalty0 1145--1159,
  1997.

\bibitem[Breiman(1996)]{bagging}
Breiman, L.
\newblock Bagging predictors.
\newblock \emph{Machine learning}, 24\penalty0 (2):\penalty0 123--140, 1996.

\bibitem[Breiman(2001)]{forest}
Breiman, L.
\newblock Random forests.
\newblock \emph{Machine learning}, 45\penalty0 (1):\penalty0 5--32, 2001.

\bibitem[Breiman et~al.(1984)Breiman, Friedman, Olshen, and Stone]{cart}
Breiman, L., Friedman, J., Olshen, R.~A., and Stone, C.~J.
\newblock \emph{Classification and regression trees}.
\newblock CRC press, 1984.

\bibitem[Brier(1950)]{brier}
Brier, G.~W.
\newblock Verification of forecasts expressed in terms of probability.
\newblock \emph{Monthly weather review}, 78\penalty0 (1):\penalty0 1--3, 1950.

\bibitem[Cappelli et~al.(2002)Cappelli, Mola, and Siciliano]{cappelli}
Cappelli, C., Mola, F., and Siciliano, R.
\newblock A statistical approach to growing a reliable honest tree.
\newblock \emph{Computational statistics \& data analysis}, 38\penalty0
  (3):\penalty0 285--299, 2002.

\bibitem[Celebi et~al.(2007)Celebi, Kingravi, Uddin, Iyatomi, Aslandogan,
  Stoecker, and Moss]{celebi}
Celebi, M.~E., Kingravi, H.~A., Uddin, B., Iyatomi, H., Aslandogan, Y.~A.,
  Stoecker, W.~V., and Moss, R.~H.
\newblock A methodological approach to the classification of dermoscopy images.
\newblock \emph{Computerized Medical Imaging and Graphics}, 31\penalty0
  (6):\penalty0 362--373, 2007.

\bibitem[Cozza et~al.(2011)Cozza, Guarracino, Maddalena, and Baroni]{cozza}
Cozza, V., Guarracino, M.~R., Maddalena, L., and Baroni, A.
\newblock Dynamic clustering detection through multi-valued descriptors of
  dermoscopic images.
\newblock \emph{Statistics in medicine}, 30\penalty0 (20):\penalty0 2536--2550,
  2011.

\bibitem[D'Ambrosio et~al.(2012)D'Ambrosio, Aria, and Siciliano]{dambrosio}
D'Ambrosio, A., Aria, M., and Siciliano, R.
\newblock Accurate tree-based missing data imputation and data fusion within
  the statistical learning paradigm.
\newblock \emph{Journal of classification}, 29\penalty0 (2):\penalty0 227--258,
  2012.

\bibitem[Dietterich(2000)]{dietterich}
Dietterich, T.~G.
\newblock Ensemble methods in machine learning.
\newblock In \emph{Multiple classifier systems}, pages 1--15. Springer, 2000.

\bibitem[Ferri et~al.(2009)Ferri, Hern{\'a}ndez-Orallo, and Modroiu]{ferri}
Ferri, C., Hern{\'a}ndez-Orallo, J., and Modroiu, R.
\newblock An experimental comparison of performance measures for
  classification.
\newblock \emph{Pattern Recognition Letters}, 30\penalty0 (1):\penalty0 27--38,
  2009.

\bibitem[Freund and Schapire(1997)]{boosting}
Freund, Y. and Schapire, R.~E.
\newblock A decision-theoretic generalization of on-line learning and an
  application to boosting.
\newblock \emph{Journal of computer and system sciences}, 55\penalty0
  (1):\penalty0 119--139, 1997.

\bibitem[Hastie et~al.(2005)Hastie, Tibshirani, Friedman, and
  Franklin]{statlearn}
Hastie, T., Tibshirani, R., Friedman, J., and Franklin, J.
\newblock The elements of statistical learning: data mining, inference and
  prediction.
\newblock \emph{The Mathematical Intelligencer}, 27\penalty0 (2):\penalty0
  83--85, 2005.

\bibitem[Kruskal and Wallis(1952)]{kwanova}
Kruskal, W. H., and Wallis, W. A. 
\newblock Use of ranks in one-criterion variance analysis. 
\newblock \emph{Journal of the American statistical Association}, 47\penalty0 (260): \penalty0 583--621, 1952


\bibitem[Limam et~al.(2003)Limam, Diday, and Winsberg]{limam}
Limam, M., Diday, E., and Winsberg, S.
\newblock Symbolic class description with interval data.
\newblock \emph{Journal of Symbolic Data Analysis}, 1\penalty0 (1), 2003.

\bibitem[Maglogiannis and Kosmopoulos(2006)]{maglogiannis}
Maglogiannis, I. and Kosmopoulos, D.~I.
\newblock Computational vision systems for the detection of malignant melanoma.
\newblock \emph{Oncology reports}, 15\penalty0 (4):\penalty0 1027--1032, 2006.

\bibitem[Mballo and Diday(2005)]{mballo2}
Mballo, C. and Diday, E.
\newblock Decision trees on interval valued variables.
\newblock \emph{The electronic journal of symbolic data analysis}, 3\penalty0
  (1):\penalty0 8--18, 2005.

\bibitem[Nachbar et~al.(1994)Nachbar, Stolz, Merkle, Cognetta, Vogt,
  Landthaler, Bilek, Braun-Falco, and Plewig]{nachbar}
Nachbar, F., Stolz, W., Merkle, T., Cognetta, A.~B., Vogt, T., Landthaler, M.,
  Bilek, P., Braun-Falco, O., and Plewig, G.
\newblock The abcd rule of dermatoscopy: high prospective value in the
  diagnosis of doubtful melanocytic skin lesions.
\newblock \emph{Journal of the American Academy of Dermatology}, 30\penalty0
  (4):\penalty0 551--559, 1994.

\bibitem[Otsu(1975)]{otsu}
Otsu, N.
\newblock A threshold selection method from gray-level histograms.
\newblock \emph{Automatica}, 11\penalty0 (285-296):\penalty0 23--27, 1975.

\bibitem[Siciliano et~al.(2004)Siciliano, Aria, and Conversano]{harvesting}
Siciliano, R., Aria, M., and Conversano, C.
\newblock Harvesting trees: methods, software and applications.
\newblock In \emph{Proceedings in Computational Statistics: 16th Symposium of
  IASC. COMPSTAT2004, held Prague}, 2004.

\bibitem[Siciliano et~al.(2010)Siciliano, Tutore, Aria, and
  D'Ambrosio]{sicilianotrees}
Siciliano, R., Tutore, V.~A., Aria, M., and D'Ambrosio, A.
\newblock Trees with leaves and without leaves.
\newblock In \emph{45th scientiï¬c meeting of the Italian Statistical
  Society}. Italian Statistical Society, 2010.

\bibitem[Tutore et~al.(2007)Tutore, Siciliano, and Aria]{tutore}
Tutore, V.~A., Siciliano, R., and Aria, M.
\newblock Conditional classification trees using instrumental variables.
\newblock In \emph{Advances in Intelligent Data Analysis VII}, pages 163--173.
  Springer, 2007.

\bibitem[Zadrozny and Elkan(2001)]{zadrozny}
Zadrozny, B. and Elkan, C.
\newblock Obtaining calibrated probability estimates from decision trees and
  naive bayesian classifiers.
\newblock In \emph{ICML}, volume~1, pages 609--616. Citeseer, 2001.

\end{thebibliography}
\end{document}